\documentclass[sigconf]{acmart}

\def\BibTeX{{\rm B\kern-.05em{\sc i\kern-.025em b}\kern-.08emT\kern-.1667em\lower.7ex\hbox{E}\kern-.125emX}}

\usepackage{subfig}
\usepackage[ruled]{algorithm2e}

\copyrightyear{2021}
\acmYear{2021}
\setcopyright{acmcopyright}\acmConference[ACSAC '21]{Annual Computer Security Applications Conference}{December 6--10, 2021}{Virtual Event, USA}
\acmBooktitle{Annual Computer Security Applications Conference (ACSAC '21), December 6--10, 2021, Virtual Event, USA}
\acmPrice{15.00}
\acmDOI{10.1145/3485832.3485900}
\acmISBN{978-1-4503-8579-4/21/12}

\begin{document}

\title{ReCFA: Resilient Control-Flow Attestation}

\author{Yumei Zhang}
\authornote{Both authors contributed equally to this research and are co-first authors.}
\email{zhangyumei319@163.com}
\author{Xinzhi Liu}
\authornotemark[1]
\email{1144518670@qq.com}
\affiliation{%
  \institution{Xidian University}
  \city{Xi'an}
  \country{China}
  \postcode{710071}
}

\author{Cong Sun}
\authornote{Corresponding author}
\email{suncong@xidian.edu.cn}
\orcid{0000-0001-9116-2694}
\affiliation{%
  \institution{Xidian University}
  \city{Xi'an}
  \country{China}
  \postcode{710071}
}

\author{Dongrui Zeng}
\email{dxz16@psu.edu}
\affiliation{%
  \institution{Pennsylvania State University}
  \city{University Park}
  \state{PA}
  \country{USA}
}

\author{Gang Tan}
\email{gtan@psu.edu}
\affiliation{%
  \institution{Pennsylvania State University}
  \city{University Park}
  \state{PA}
  \country{USA}
}

\author{Xiao Kan}
\email{814091656@qq.com}
\affiliation{%
  \institution{Xidian University}
  \city{Xi'an}
  \country{China}
  \postcode{710071}
}

\author{Siqi Ma}
\email{slivia.ma@uq.edu.au}
\affiliation{%
  \institution{The University of Queensland}
  \city{Brisbane}
  \country{Australia}
}

\begin{abstract}
Recent IoT applications gradually adapt more complicated end systems with commodity software. Ensuring the runtime integrity of these software is a challenging task for the remote controller or cloud services. Popular enforcement is the runtime remote attestation which requires the end system (prover) to generate evidence for its runtime behavior and a remote trusted verifier to attest the evidence. Control-flow attestation is a kind of runtime attestation that provides diagnoses towards the remote control-flow hijacking at the prover. Most of these attestation approaches focus on small or embedded software. The recent advance to attesting complicated software depends on the source code and CFG traversing to measure the checkpoint-separated subpaths, which may be unavailable for commodity software and cause possible context missing between consecutive subpaths in the measurements.

In this work, we propose a resilient control-flow attestation (ReCFA), which does not need the offline measurement of all legitimate control-flow paths, thus scalable to be used on complicated commodity software. Our main contribution is a multi-phase approach to condensing the runtime control-flow events; as a result, the vast amount of control-flow events are abstracted into a deliverable size. The condensing approach consists of filtering skippable call sites, folding program-structure related control-flow events, and a greedy compression. Our approach is implemented with binary-level static analysis and instrumentation. We employ a shadow stack mechanism at the verifier to enforce context-sensitive control-flow integrity and diagnose the compromised control-flow events violating the security policy. The experimental results on real-world benchmarks show both the efficiency of the control-flow condensing and the effectiveness of security enforcement.
\end{abstract}



\maketitle

\section{Introduction}

The tampering of the software implementation as well as its runtime state on devices poses a critical challenge for the security of IoT. To mitigate such vulnerabilities, remote attestation has been widely deployed as a security service to measure the integrity of software on resource-constrained end devices. Based on some interactive security protocol, the remote device, i.e. \emph{prover}, sends an authentication report about its software status to a trusted party, i.e. \emph{verifier}, to prove that it has not been tampered with or hijacked. The freshness and authenticity of the report are usually ensured by a \emph{trust anchor} on the prover. The instances of trust anchors vary from the heavyweight Trusted Platform Module (TPM) to more lightweight schemes such as TrustLite\cite{DBLP:conf/eurosys/KoeberlSSV14}, Sancus\cite{DBLP:conf/uss/NoormanADSHHPVP13}, and ARM TrustZone-M\cite{arm}.

The static remote attestation schemes verify the integrity of program code, executables, and the configurations on the prover. They cannot capture the runtime compromises related to memory errors, e.g., the control-flow hijacking based on code reuse \cite{DBLP:conf/ccs/Shacham07, DBLP:conf/ccs/BuchananRSS08}. On the other hand, the runtime remote attestations~\cite{DBLP:conf/ccs/DaviSW09, DBLP:conf/ccs/AberaADENPST16, DBLP:conf/ndss/AberaBB0SS19, DBLP:conf/raid/ToffaliniLB0C19, DBLP:conf/iccad/ZeitouniDAS0JS17, DBLP:conf/dac/DessoukyZNPDKAS17, DBLP:conf/iccad/DessoukyA0S18, DBLP:journals/corr/abs-2011-07400, DBLP:conf/sp/SunFLJ20} attempt to measure the runtime behaviors of software at the prover and quote fine-grained runtime status to the verifier. For example, some approaches~\cite{DBLP:conf/ccs/AberaADENPST16, DBLP:conf/raid/ToffaliniLB0C19, DBLP:conf/iccad/ZeitouniDAS0JS17, DBLP:conf/dac/DessoukyZNPDKAS17, DBLP:conf/iccad/DessoukyA0S18, DBLP:journals/corr/abs-2011-07400} implement fine-grained control-flow measurements delivered to the verifier to diagnose the concrete execution path under attack. On receiving the hashed information of the prover's execution paths to the verifier, the verifier reports control-flow hijacking or non-control data attacks by detecting the absence of the received hashes in a pre-measured database. Remotely attesting control-flow integrity (CFI) is more challenging than traditional local CFI enforcement because the verifier can only rely on the abstract measurements received from the prover, whereas local enforcement can inline the checking to use all the runtime information.

The advance of mobile edge computing \cite{DBLP:journals/comsur/MachB17} allows for more powerful end-systems with complicated commodity software, whose runtime integrity is better to be confirmed by remote management servers. Although the control-flow attestations have shown an advantage on remotely detecting control-flow hijacking, most of them fall short of scalability on complicated software. More specifically, the control-flow attestations rely on an offline procedure to measure all the legitimate control-flow paths, which is potentially exponential to the scale of the program. Such an offline procedure, though feasible on IoT programs, is unrealistic to large programs or services. ScaRR \cite{DBLP:conf/raid/ToffaliniLB0C19} is the first work that adapts the runtime attestation to complex systems. The new control-flow model of ScaRR separates the control-flow path with checkpoints and represents each subpath between consecutive checkpoints with a sequence of control-flow events, including procedure calls, returns, and branches. A remote shadow stack at the verifier is used to preserve the calling context for precision. However, there are still limitations to this approach. ScaRR still requires the offline procedure to obtain a mapping from the measurement of each subpath to a sequence of the critical control-flow events on that subpath. To make the subpaths and measurements correct, source code is required, which is unobtainable for commodity software. Also, measuring the checkpoints-separated subpaths may cause context missing between consecutive subpaths to bring in potential false negatives of the online remote verification. Last, the verification of ScaRR only provides coarse-grained path diagnoses. ScaRR can identify subpaths triggering control-flow violations but fails to locate the exact control-flow events (e.g. indirect branches) that cause the violations.


To mitigate such limitations, in this work, we propose a new binary-level \emph{resilient control-flow attestation} (ReCFA) with no offline generation of path measurements. The executable of the prover is statically instrumented to record critical runtime control-flow events, which are further condensed at runtime for efficient delivery to the verifier. The verifier recovers the control-flow events and monitors the control-flow integrity with a shadow stack. To overcome the runtime explosion of control-flow events, we propose a multi-phase control-flow event condensing approach, which folds the loops and recursions at runtime. Our approach is resilient in two aspects. First, due to the binary-level analysis and the avoidance of offline exponential paths measuring, our approach can verify the runtime behaviors of complicated commodity programs. Second, our approach can enforce different CFI policies specified by different binary-level CFGs, e.g. \cite{DBLP:conf/sp/VeenGCPCRBHAG16,DBLP:conf/codaspy/ZengT18,DBLP:conf/ndss/KimSZT21}. We summarize our contributions as follows:
\begin{enumerate}
\item We propose a new control-flow attestation for large binary programs. The approach avoids offline path measurements generation and can enforce fine-grained context-sensitive CFI with remote shadow stack.

\item We propose a novel multi-phase condensing and recovering approach for the control-flow events to efficiently encode, deliver, and attest the runtime behavior of the prover.

\item Our evaluation results show the efficiency and security of ReCFA. The attestation speed is around 28.2M/s and the verification speed is around 1.03M/s, which justify the usability of our approach under various wireless environments.

\end{enumerate}

\section{Design of ReCFA}

ReCFA is a control-flow attestation framework using static binary analysis and binary instrumentation to enforce remote CFI and diagnose control-flow hijacking remotely for complex software. We first outline its architecture, the threat model, and the requirements under which ReCFA operates. Then, we present our efficient control-flow event abstraction and security enforcement.

\subsection{System Overview}

\begin{figure}[!ht]
\centering
\includegraphics[width=3.3in]{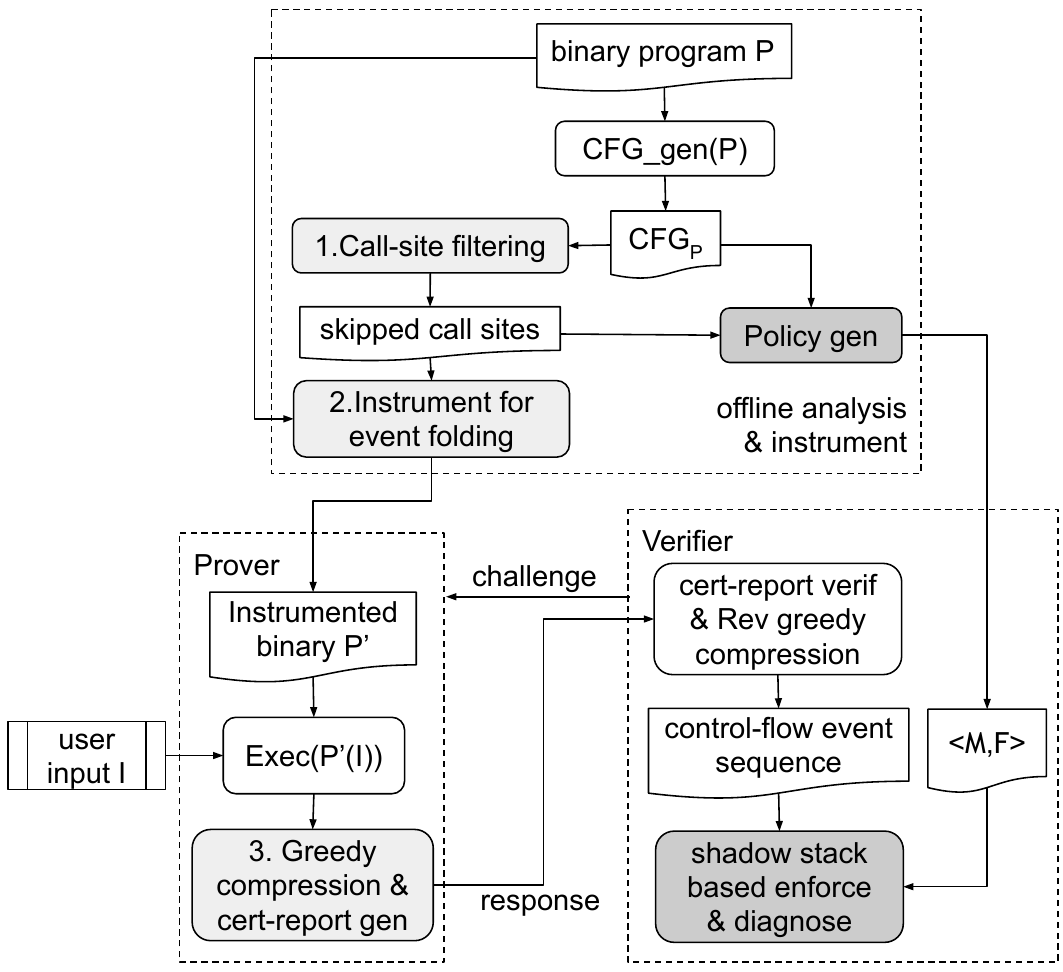}
\caption{Architecture of ReCFA}\label{fig:arch}
\end{figure}

The architecture of ReCFA is presented in Fig.~\ref{fig:arch}. ReCFA first relies on an offline static analysis and instrumentation procedure to generate the prover program and the policies deployed on the verifier. In this stage, we first derive a binary-level control-flow graph of the program. Then we filter the skippable direct calls (Section~\ref{subsec:call-site-filtering}) and instrument the functionality of runtime control-flow event folding (Section~\ref{subsec:folding}). We derive the security policy in the form of two mappings, one for tracking the skipped edges and the other for the policy enforcement (Section~\ref{subsec:enforcement}). The instrumented program running at the prover generates folded control-flow event sequences, which are compressed with a greedy algorithm (Section~\ref{subsec:greedy}) and sent to the verifier as the attestation report. The verifier recovers control-flow event sequences to attest against the CFI policy with the shadow stack mechanism (Section~\ref{subsec:enforcement}). Any event sequence violating the CFI policy leads to a remote diagnosis of the prover program's vulnerable control flows.

\subsection{Threat Model and Requirements}

As an attestation scheme, our attacker model should include the network threats (e.g., replay or masquerading attacks) to the traditional remote attestation schemes. However, ensuring the freshness and authenticity of our attestation report between the prover and verifier can rely on the state-of-the-art attestation protocol and a trust anchor. Though several existing trust anchors, e.g., \cite{DBLP:conf/uss/NoormanADSHHPVP13, DBLP:journals/iacr/CostanD16}, are available, we assume the kernel combined with the hardware-assisted protection keys (MPK) \cite{mpk} as the trust anchor of the prover, which is reasonable against user-level attackers and for commodity hardware. The trust anchor is responsible for the code integrity through static remote attestation. It is also critical to ensure the prover-side instrumented code cannot be bypassed and the protection of memory storing critical data structures.

At the system level of the prover, to follow a similar assumption as \cite{DBLP:conf/ccs/AberaADENPST16,DBLP:conf/iccad/DessoukyA0S18,DBLP:conf/raid/ToffaliniLB0C19}, the data execution prevention (DEP) is deployed on the prover to prevent malicious code injection into the running processes. We focus on the control-flow hijacking on the prover. The attacker can run the program with arbitrary input, read/write the data section of the program, and exploit memory corruption vulnerabilities (e.g., buffer overflow) to manipulate the in-memory control-flow information and hijack the program's control flow. Though pursuing a similar threat model as CFI, the attestation scheme requires the verifier to remotely diagnose the control-flow path leading to the control-flow attack. This requirement is unsatisfied by the traditional local CFI protections.

Besides, the attacker model does not include physical attacks or data-oriented attacks that do not alter the control-flow edges. The attacker cannot emit self-modifying code, runtime generated code, or the unanticipated dynamic loading of code \cite{DBLP:journals/tissec/AbadiBEL09}.

\subsection{Multi-Phase Control Flows Condensing}\label{sec:condensing}

In this section, we propose an approach to reducing the overall amount of data delivered from the prover to the verifier. For the context-sensitive control-flow attestation with shadow stack, these data include all the function calls, indirect jumps, and returns encountered at runtime by the prover program, which we define as the \emph{Potential Monitoring Points} (PMPs).

\subsubsection{Phase-1: Call-Site Filtering}\label{subsec:call-site-filtering}

Recording all the function calls by instrumentation causes remarkable performance overhead. Thus, our first step is to filter out call sites that are unnecessary for instrumentation. We use the causality relation between the consecutive PMPs to identify \emph{Skippable Call Sites} (SCSes).

Specifically, we build an abstract graph $G=(V,E)$ for each program to discover SCSes, where the node set $V$ contains all the PMPs and function entry addresses for the functions that no call site targets at. If for any $v, v'\in V$, we find a control-flow path from $v$ to $v'$ without any intermediate $v''\in V$, then we add a directed edge $(v,v')$ to $E$. With this abstract graph $G$, the principle of skipping certain node (e.g., direct calls) is as follows.
A node is skippable only when none of its predecessors has more than one successor. Intuitively, we skip a node when the node is guaranteed to be executed subsequently after each of its predecessor.

\begin{figure}

\subfloat[Code Sample]{
  \includegraphics[width=\linewidth]{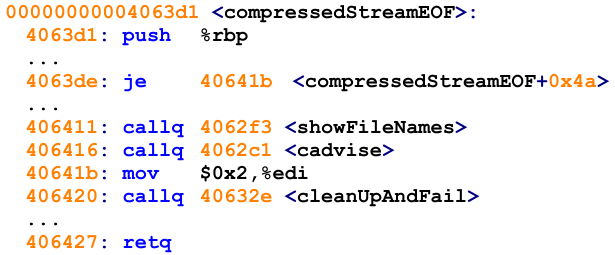}
  \label{fig:filter-program}
}

\subfloat[Abstract Graph]{
  \includegraphics[width=\linewidth]{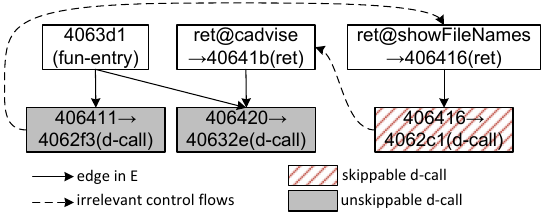}
  \label{fig:filter-graph}
}

\caption{Example of Filtered Call Site}
\end{figure}

Taking the program in Fig.~\ref{fig:filter-program} for example, the corresponding abstract graph is presented in Fig.~\ref{fig:filter-graph}. The abstract graph is different from CFG. Considering the two nodes representing the return edges of CFG in Fig.~\ref{fig:filter-graph}, they contain the return target information, i.e., 40641b and 406416. Even this \verb|ret| instruction in \verb|showFileNames| may return to different targets, there is only one CFG edge (targeting at 406416) can be the predecessor of the direct-call node ``406416$\rightarrow$4062c1''.
To explain which direct-call node is skippable, for the call at 406416, since it has only one predecessor, i.e., the return node of function \verb|showFileNames| to 406416, and this return node has only one successor, the call at 406416 to function \verb|cadvise| is skippable. For the call at 406420 to function \verb|cleanUpAndFail|, it has two predecessors. One is the return node from function \verb|cadvise| to 40641b. The other is the entry 4063d1 of function \verb|compressedStreamEOF| bridged by the conditional jump at 4063de. Because the entry 4063d1 has two successors (the call at 406411 and 406420), the call at 406420 cannot be skipped.

The skipped call sites derived by this procedure should also be held by the verifier to enforce the security policy with the shadow stack. For the skipped call-site nodes $S\subseteq V$, we find all the predecessor nodes $P\subseteq V$ in $G$. We build a mapping $\mathcal{M}$ from the predecessors' target addresses to the skipped call sites, i.e., $\mathcal{M}=\{t\mapsto s'\mid (s,t)\in P \wedge (s',t')\in S \wedge ((s,t),(s',t'))\in E\}$. For the example in Fig.~\ref{fig:filter-graph}, we know $(406416,4062c1)\in S$ and $(\verb|ret@showFileNames|,$ $406416)\in P$, then we add $\{406416\mapsto 406416\}$ into $\mathcal{M}$. An exceptional case is for the direct-call predecessors, we encode their call-site address as the key in the mapping, i.e. adding $\{s\mapsto s'\}$ into $\mathcal{M}$ instead of $\{t\mapsto s'\}$. We elaborate the reason in the control-flow edge encoding of Section~\ref{sec:implementation}.

\subsubsection{Phase-2: Control-Flow Events Folding}\label{subsec:folding}

The requirement of diagnosing all critical control-flow events causes runtime explosion of data sent to the verifier. The explosion is mainly caused by the loops and recursions of programs. To achieve a reasonable communication overhead, we design a folding mechanism to efficiently capture the unskipped control-flow events in loops and recursions.

\begin{figure}

\subfloat[Loop Example]{
  \includegraphics[width=\linewidth]{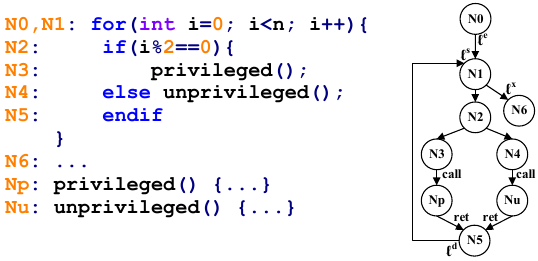}
  \label{fig:sub-loop}
}

\subfloat[Recursion Example]{
  \includegraphics[width=\linewidth]{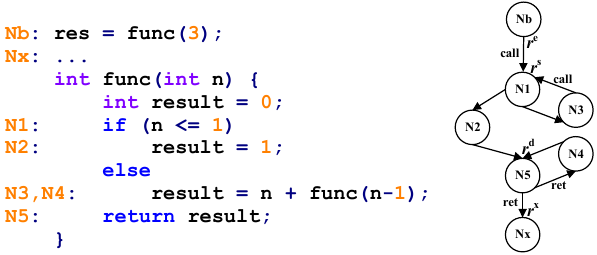}
  \label{fig:sub-rec}
}

\subfloat[Excluded Recursion]{
  \includegraphics[width=\linewidth]{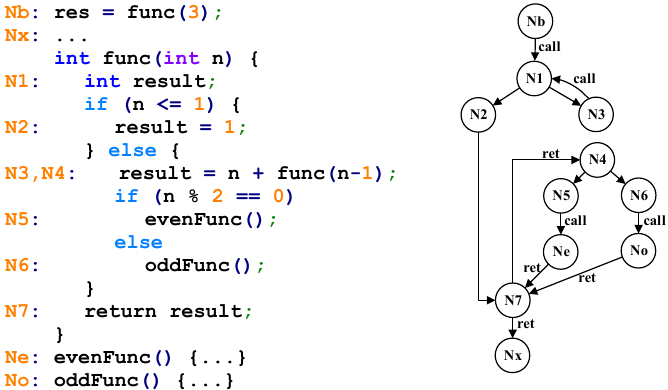}
  \label{fig:sub-rec-cond}
}

\caption{Program Examples}
\end{figure}

For each loop $\ell$ in the binary, we identify four kinds of points for the binary instrumentation, i.e., loop entry ($\ell^e$), loop exit ($\ell^x$), loop body start ($\ell^s$), and loop body end ($\ell^d$). An effective static analysis to identify these points requires the binary program to have structured control flows, which is common for open source software and assumed by other work using binary rewriting, e.g., \cite{DBLP:conf/ccs/AberaADENPST16}. However, this doesn't hold for arbitrary or obfuscated binaries. In our case, the control-flow events are directly fed into the greedy compression procedure presented in Section~\ref{subsec:greedy}, which treats the control-flow events as a stream. Thus, ReCFA's runtime folding mechanism supports obfuscated binaries. For the example in Fig.~\ref{fig:sub-loop}, we label the critical points in the control-flow graph. We instrument code at these points to conduct the control-flow event folding.

\begin{figure}[!t]
\centering
\includegraphics[width=3.3in]{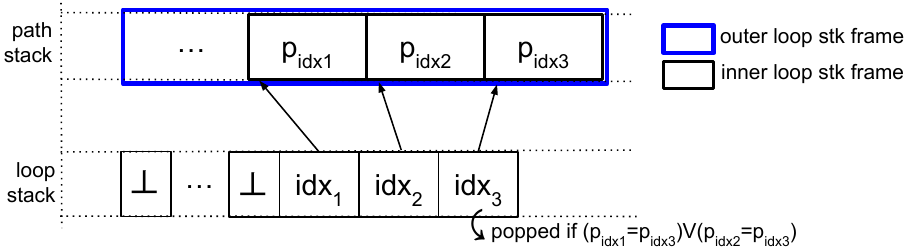}
\caption{Folding Nested Loops}\label{fig:principle}
\end{figure}

To fold the loop, we define two runtime data structures: \emph{loop stack} and \emph{path stack}, as presented in Fig.~\ref{fig:principle}. The loop stack stores the status of the ongoing (unclosed) and maybe nested loops. If the program counter is out of any loop, the loop stack should be empty. Each element of the loop stack is an index of the path stack, except several special tag $\bot$ to demarcate the inner and outer loop. Each index represents an id of a \emph{stack frame} on the path stack. The path stack consists of a number of stack frames. Each stack frame holds the control-flow events (calls, returns, indirect jumps) as deduplicated \emph{event paths} captured during the execution of a specific loop. The stack frames of an inner loop can be nested into a stack frame of an outer loop.

Considering the program in Fig.~\ref{fig:sub-loop}, when the program counter reaches a loop entry, i.e., the last instruction of $N_0$, we instrument to push a tag $\bot$ onto the loop stack. At the start of a loop iteration, i.e., the first instruction of $N_1$, we label the top of the path stack as the start of a new stack frame for a new event path, and push the top index of the path stack onto the loop stack. During the execution of the current iteration, the control-flow events accumulate into the new stack frame as an event path, i.e., $p_{idx_1}=\{(N_3,N_p),(N_p,N_5)\}$. When this iteration reaches an end at the last instruction of $N_5$, we compare the top event path $p_{idx_1}$ with the event paths indexed by the elements of the loop stack above the top-most $\bot$. Because no other event path is indexed above the top-most $\bot$, we have not found path duplication at this time and $p_{idx_1}$ is reserved. Then the second iteration starts from $N_1$ to $N_5$ and $p_{idx_2}=\{(N_4,N_u),(N_u,N_5)\}$ is accumulated on the path stack. $p_{idx_2}$ is also reserved because it is a different event path compared with $p_{idx_1}$. From the third iteration, the accumulated event path $p_{idx_3}$ should be identical to either $p_{idx_1}$ or $p_{idx_2}$. Since duplicated event path is found, we pop the top event path $p_{idx_3}$ and its index $idx_3$ from the path stack and loop stack respectively. Finally, when the loop ends at the first instruction of $N_6$, we pop the content of the loop stack above the top-most $\bot$. The two deduplicated event paths $\{(N_3,N_p),(N_p,N_5)\}$ and $\{(N_4,N_u),(N_u,N_5)\}$ become a part of the outer stack frame to continue the folding of the outer loop.

To fold the recursions, we firstly detect four kinds of points, the start and return points of recursion function ($r^s$, $r^d$), the external call sites to recursive functions ($r^e$), and the external call-after points of recursive functions ($r^x$), which stand for the end of recursion. Considering the recursion program in Fig.~\ref{fig:sub-rec}, at the external call site to the function \verb|func|, i.e. the last instruction of $N_b$, we push a tag $\bot$ onto the \emph{recursion stack} which is an alias of loop stack here. At the start and return points ($r^s$, $r^d$) of \verb|func|, we instrument deduplication actions to check if the currently accumulated top event path (if exists) is identical to some other event path indexed by the elements of the recursion stack above the top-most $\bot$. If duplicated event paths are found, the top event path and its index are popped from the path stack and recursion stack respectively. In consequence, after pushing $\{(N_b,N_1)\}$ on the path stack, $\{(N_3,N_1)\}$ and $\{(N_5, N_4)\}$ are pushed and deduplicated for many times. At the end of the recursion, i.e. the start of $N_x$, another event path $\{(N_5, N_x)\}$ is pushed and we pop the content of the recursion stack above the top-most $\bot$.

The loop stack and recursion stack are implemented with the same data structure. This design can naturally deal with the interleaving loops and recursions. There are recursion cases that cause the above recursion folding approach to trigger false positives at the verifier. Considering the recursion in Fig.~\ref{fig:sub-rec-cond}, when using the above recursion folding, we derive a path $\{(N_b,N_1),(N_3,N_1), (N_7,N_4), $ $(N_5,N_e), (N_e,N_7), (N_7,N_4), (N_6,N_o), (N_o,N_7), (N_7,N_x)\}$. This path contains four calls and five returns, thus will trigger a false alert of CFI violation at the verifier. The reason is that from $N_4$ to $N_7$ there are two subpaths of control-flow events, making the outer recursive call $(N_3,N_1)$ fail to match with the two $(N_7,N_4)$. To avoid these false positives, we use static analysis to detect if there are multiple control-flow event paths from $N_4$ to $N_7$. The analysis is an offline depth-limited depth-first traversal. If there is only one path from $N_4$ to $N_7$, we apply the recursion folding; otherwise, we treat the points $N_b, N_1, N_7$, and $N_x$ as the ordinary points of control-flow events to avoid folding this recursion.

\subsubsection{Phase-3: Greedy Compression}\label{subsec:greedy}

The above recursion folding requires static analysis to correctly locate the critical points for instrumentation, which prevents our approach from being applied on mutual recursion. We propose a path compression algorithm to further reduce the size of control-flow events to be sent to the verifier. The algorithm takes the sequence of control-flow events generated by Phase-2 as input. The compression will insert several \emph{leading knots} for the compressed part of the sequence. For example, if the control-flow event sequence is $e_1 e_2 e_3 e_4 e_2 e_3 e_4 e_5$, the output of the algorithm will be $e_1 \langle 2,3\rangle e_2 e_3 e_4 e_5$ where the knot $\langle 2,3\rangle$ means the subsequent events with length 3 are repeated twice. The knots are encoded as a number distinguishable from code addresses.

\begin{algorithm}[t]

$idx \gets 0; r\gets []$\;
\For{$pos_w \gets 0$ \KwTo $length(p) - 1$}{
    $n_{rep} \gets 0$; $sz_w \gets 1$\;
    \While{$sz_w < BOUND$}{
        $pos_{chk} \gets pos_w + sz_w*(n_{rep}+1)$\;
        \If{$pos_{chk} + sz_w > length(p) \wedge n_{rep}=0$}{
            \textbf{break}\;
        }
        \For{$j\gets 0$ \KwTo $sz_w \wedge pos_{chk} + j < length(p)$}{
            \If{$ p[pos_w+j]\neq p[pos_{chk}+j]$}{
                \textbf{break}\;
            }
        }
        \uIf{$j=sz_w$}{
            $n_{rep}\gets n_{rep}+1$\;
        }
        \uElseIf{$n_{rep} = 0$}{
            $sz_w\gets sz_w +1$\;
        }
        \Else{
            $knot(r, idx, \langle n_{rep}+1, sz_w\rangle)$\;
            $r[idx .. (idx+sz_w)] \gets p[pos_w .. (pos_w+sz_w)]$\;
            $idx \gets idx + sz_w$\;
            $pos_w \gets pos_w + sz_w*(n_{rep}+1)$\;
            $n_{rep}\gets 0$; $sz_w \gets 1$\;
        }
    }
    $r[idx] \gets p[pos_w]$\;
    $idx \gets idx +1$\;
}
\textit{compress}($r,idx$)\;

\caption{GreedyCompression($p,BOUND$)}\label{algo:greedy}
\end{algorithm}

The compression algorithm is presented in Algorithm~\ref{algo:greedy}. $p$ is the input control-flow event sequence. $r$ is the list to store the compression result. $BOUND$ is the upper bound of a sliding window. The sliding window has a starting position $pos_w$ and a size $sz_w$. $n_{rep}$ holds the times of repetition of the content in the sliding window. When the content in the sliding window is found repeated, the algorithm iterates on the while-loop to capture the repeating times into $n_{rep}$. The predicate $knot(r,pos,k)$ means attaching knot number $k$ at the position $pos$ of the list $r$. The final predicate $compress(l,n)$ means compressing the size-$n$ content in the list $l$, as well as the attached knot numbers, with an off-the-shelf compression algorithm, Zstandard \cite{zstandard} as reported to outperform other compressions in \cite{DBLP:conf/raid/ToffaliniLB0C19}. We choose the greedy algorithm instead of migrating from the Rabin-Karp rolling hashing because hash-based comparison may introduce false positives. The complexity of the algorithm is $O(n\times BOUND)$. Also, the greedy matching tends to fold short repeated subsequences instead of the longest ones for efficiency. For example, the event sequence $e_1 e_2 e_1 e_2 e_3 e_1 e_2 e_1 e_2 e_3$ will be compressed to $\langle 2,2\rangle e_1 e_2 e_3 \langle 2,2\rangle e_1 e_2 e_3$ instead of $\langle 2,5\rangle e_1 e_2 e_1 e_2 e_3$. A later step is to generate the key-based authentication code and nonce for the compression results used in the procedure of ordinary remote attestation, which is out of the scope of this work.

\subsection{Context-Sensitive Remote Enforcement}\label{subsec:enforcement}

The context-sensitive enforcement on the verifier relies on a shadow stack. Traditional shadow stacks are located on the monitored address space and need to be protected in safe memory regions, e.g., by segmentation with hardware or isolation with SFI. In contrast, our shadow stack is held at the verifier side which is assumed under protection and resisting the attack like \cite{DBLP:conf/sp/EvansFGOTSSRO15}.


Different from using the CFG for generating the measurements \cite{DBLP:conf/raid/ToffaliniLB0C19}, we use static binary-level arity-based CFG generated by TypeArmor \cite{DBLP:conf/sp/VeenGCPCRBHAG16} directly as our security policy.
The indirect jumps are resolved by Dyninst \cite{DBLP:conf/paste/BernatM11}. Because we only monitor function calls, indirect jumps, and returns at the prover, the control-flow edges other than these edges are unnecessary for the verifier. To facilitate the enforcement, we derive another mapping $\mathcal{F}$ statically for the forward edges. The element of $\mathcal{F}$ is of form $cs\mapsto (ca,tgts)$. The key $cs$ is the call site address of a forward edge; $ca$ is the address of the call-after point of the call site and $tgts$ is the set of valid target addresses of the call. Different types of forward edges can be identified in the mapping. For an indirect jump at address $addr$, we encode it as $addr\mapsto (\bot, jmp\_tgts)$. For a direct call at $cs$, we encode it as $cs\mapsto (ca,\emptyset)$ because the target of a direct call can not be counterfeited by our attacker model thus not monitored at the prover.

The first step of the runtime enforcement is to recover the control-flow event sequence by a reverse procedure of Phase-3 discussed in Section~\ref{sec:condensing}, which is trivial due to the knot numbers. Then, for each edge $(s,t)$ in the recovered control-flow event sequence, we look up $s$ in the mapping $\mathcal{F}$ to decide if the edge is a forward edge. If it is a direct call, we push the call-after point tracked from $\mathcal{F}$ onto the shadow stack. If it is an indirect jump, we check if its runtime target $t$ is in the $jmp\_tgts$ tracked from $\mathcal{F}$ and report security violation when $t\notin jmp\_tgts$. If it is an indirect call, after confirming its runtime target $t$ is in the valid targets tracked from $\mathcal{F}$, we push its call-after point onto the shadow stack. For the backward edge, we check if its return target is identical to the top element of the shadow stack.

A critical step for the correct enforcement is to recover the skipped direct call sites from the control-flow event sequence. The procedure is, for each consecutive event pair, e.g., $e_i e_{i+1}$ in the sequence, we recursively track the mapping $\mathcal{M}$ initially from $e_i$ to find the subsequent skipped edge sequence of $e_i$ that reaches $e_{i+1}$. We apply the above enforcement strategies on these skipped edges before the edge $e_{i+1}$. For the example in Fig.~\ref{fig:filter-graph}, after the enforcement dealing with the return to 406416 (by matching 406416 with and then popping the top of shadow stack), we look up 406416's value in the mapping $\mathcal{M}$; in this case, 406416. It means the skipped call at 406416 should be dealt with at once by tracking 406416 at the mapping $\mathcal{F}$.

The \verb|setjmp/longjmp| breaks the calling convention to enable non-local indirect jumps and challenges the shadow stack's tracking. To deal with this case, when we find the top expected return target on the shadow stack mismatching the runtime return target from the event sequence, we pop more elements iteratively from the shadow stack for comparison until some element of shadow stack matches the runtime return target, or the shadow stack becomes empty. The latter case accounts for the control-flow hijacking.

\section{Implementation}\label{sec:implementation}

In this section, we elaborate implementation details of our approach.

\paragraph{Static analysis}

Before the binary instrumentation, we conduct static analysis on the program binary of the prover. The analysis is still offline but is more efficient than measuring all the legitimate control-flow paths as proposed in \cite{DBLP:conf/ccs/AberaADENPST16}. First, we derive a binary-level CFG with TypeArmor \cite{DBLP:conf/sp/VeenGCPCRBHAG16}. We use this CFG to conduct the call-site filtering of Section~\ref{subsec:call-site-filtering} and identify the list of skipped call sites. Since our approach does not instrument the calls to the library functions, we skip these calls before constructing the abstract graph $G$ in the filtering procedure. Then we derive mapping $\mathcal{M}$ for the skipped call sites and mapping $\mathcal{F}$ for the forward edges. We store both mappings at the verifier. In the call-site filtering, \verb|setjmp/longjmp| should be addressed in the static analysis. \verb|longjmp| can lead to a context that calls another function. If we skip \verb|longjmp| as a library function call, the predecessor of \verb|longjmp| in $G$ may mistakenly decide its successor as skippable. Therefore, \verb|longjmp| is treated as an ordinary call in the call-site filtering. For the \verb|setjmp|, there may be many \verb|longjmp| to the subsequent instructions of \verb|setjmp| but we cannot infer them. For the direct calls successive to the subsequent instruction of \verb|setjmp|, because we cannot infer their predecessors, we conservatively treat them as unskippable.

\paragraph{Encoding control-flow edges}

The encoding of different control-flow events affects the amount of data delivered to the verifier. For the indirect calls, indirect jumps, and return, the edges are encoded as a pair of code addresses. For the unskippable direct calls, since their target cannot be hijacked by the attacker model, we only encode their call site, i.e., a single code address with a special tag bit to identify the edge type.

\paragraph{Static Binary Instrumentation}

We use Dyninst to instrument the prover program. We target at x64 binaries. The first step is to identify the specific program points for instrumenting different functionalities. For the loop and recursion points used by the control-flow events folding, we traverse the loop tree structures of CFG with Dyninst to find the loop points. We detect direct recursions by checking if a subroutine called by each function is itself. We use the pattern of instruction operators to differentiate between direct and indirect branches. For the direct calls, we only instrument the unskippable direct calls based on the results of call-site filtering. Capturing the runtime indirect branches with Dyninst is straightforward. However, for the runtime return edge, the API of Dyninst can only help to capture the return into the trampoline section after instrumentation instead of into the original code section. To recover the return target at the original code section, we pair the return instruction of the callee with the call-after point in the caller using a flip flag. The flag is set at the return instruction. At the call-after point in the caller function, we pair this point with the latest return only after confirming the flag is set. The flip flag avoids the call-after point being captured mistakenly when it is a jump target instead of a return target. Such prover-side flag-based matching is to avoid wrongly recorded control-flow events of the benign program. It does not detect the violation of backward-edge CFI, which is left to the verifier.

In the implementation, the operations of capturing and folding the control-flow events are held in and loaded from a shared object file by the instrumented code. Although the shared object can be linked to the enclave shared object and protected by SGX as the trust anchor, in this work, we use a more lightweight system-call based approach to ensure the instrumented code cannot be bypassed. To protect the data structures (loop/recursion stack and path stack) used by the control-flow events folding, we use Intel's \emph{memory protection keys} (MPK) \cite{mpk} to protect the writing of these data structures in the userspace. As we know, ScaRR \cite{DBLP:conf/raid/ToffaliniLB0C19} chooses to write the control-flow edges into the kernel space. However, we do not take this approach due to the potential enormous context switching cost for the control-flow events on an order of magnitude in gigabytes. With the system-call based approach for code integrity and MPK-based protection for data integrity, we can ensure the integrity of CFA-data collection.

First, we attach the MPK operations into the instrumented code snippets for recording the CFA data. Specifically, a code snippet first enables the writing privilege to the CFA data region, then writes data, and finally disables the writing privilege. We write the user-accessible PKRU register with instrumented system calls, which wraps the WRPKRU instruction, to switch the permission to the memory region storing the CFA data. The MPK-based switching does not provide strong memory protection since the attacker may write the PKRU register and obtain the write permission to the CFA data region. To ensure the execution of every code snippet cannot be tampered with, we insert guards at the entry and exit points of each instrumented code snippet to record the type of each snippet and the guarded point in the snippet. For example, the loop entry snippet starts with a call \textit{sys\_rec}(\textit{loop\_entry}$_{start}$) and ends with a call \textit{sys\_rec}(\textit{loop\_entry}$_{end}$). In all, a code snippet \verb|s1| looks like ``\textit{sys\_rec}(\verb|s1|$_{start}$), MPK-ops, data-ops, MPK-ops, \textit{sys\_rec}(\verb|s1|$_{end}$)''. At the kernel level, we use a kernel thread to pair the consecutive entry and exit guards with the same snipped type. The code reuses exploiting the instrumented code will cause the kernel to detect disordering and mismatching at the start- or end-point of the instrumented code snippet. For example, when an attacker hijacks control flow and jumps inside the snippet \verb|s1|, the guard \verb|s1|$_{end}$ will be executed but \verb|s1|$_{start}$ will be missing. With the guards, the only possible bypass is to hijack control flow and jump out of a snippet after the writing privilege is enabled. Then, the attacker corrupts the CFA data and jumps back to the original snippet. Our solution is to avoid indirect branches in snippets. Hence, the integrity of code snippets is guaranteed and the memory corruption attacks cannot lead to forged control-flow events.

\section{Evaluations}

Our experiments are conducted on an elastic compute service with 2.5GHz$\times$8 Intel Xeon(R) Platinum 8269CY CPU, 16GB RAM, Linux 4.15.0-135-generic kernel (Ubuntu 18.04 64-bit). To take no account of the network latency, we run the prover and verifier on the same machine. The network overhead is measured with the size of data delivered between the prover and the verifier. The binaries are compiled respectively with GCC v7.5.0 and LLVM v10.0.0 under the default optimization level. For the binary analysis and instrumentation, we use Dyninst 10.1.0. The commit id of TypeArmor we used is ee018f0. The offline procedure is on the same machine. We applied ReCFA on SPEC CPU 2006's C benchmarks. To ensure standard runtime input, we use the \verb|test| workloads to run the instrumented binaries of SPEC2k6 benchmarks.

\paragraph{RQ.1} Is the prover-side performance overhead reasonable?

We evaluate the control flows condensing procedure and reveal the effect of each phase respectively. Firstly, we evaluate the call-site reduction by the call-site filtering. We report the number of the original direct calls and the skipped direct calls in the binary-level CFG in Table~\ref{tab:call-site-filtering}. The ratio of reduction ranges 16.1\%$\sim$57.2\% for GCC binaries and 16.1\%$\sim$54.5\% for LLVM binaries. The overall reduction is around 40.5\%.

\begin{table}[!t]
\renewcommand{\arraystretch}{1.1}
\caption{Effect of Call-Site Filtering (reduction = $\Sigma\text{d-call}_\text{skipped}$ / $\Sigma\text{d-call}_\text{orig}$.)}
\label{tab:call-site-filtering}
\centering
\begin{tabular}{l|r|r|r|r}
\hline
 & \multicolumn{2}{c|}{GCC} & \multicolumn{2}{c}{LLVM} \\
 \cline{2-5}
 & \#d-call & \#d-call & \#d-call & \#d-call \\
Program & orig & skipped & orig & skipped \\
\hline
400.perlbench & 13,793 & 4,168 & 13,799 & 4,179 \\

401.bzip2 & 288 & 134 & 271 & 129 \\

403.gcc & 48,610 & 21,558 & 48,416 & 21,412 \\

429.mcf & 31 & 5 & 31 & 5 \\

433.milc & 929 & 358 & 929 & 358 \\

445.gobmk & 8,898 & 3,150 & 8,887 & 3,143 \\

456.hmmer & 2,141 & 764 & 2,141 & 764 \\

458.sjeng & 739 & 272 & 739 & 272 \\

462.libquantum & 407 & 233 & 410 & 222 \\

464.h264ref & 2,070 & 735 & 2,070 & 744 \\

470.lbm & 33 & 18 & 33 & 18 \\

482.sphinx3 & 2,064 & 1,075 & 2,064 & 1,075 \\

\hline
Overall reduction & \multicolumn{2}{c|}{40.6\%} & \multicolumn{2}{c}{40.5\%} \\
\hline
\end{tabular}
\end{table}

\begin{table*}[!t]
\renewcommand{\arraystretch}{1.1}
\caption{Effect of Control-Flow Events Folding. Overhead=$(T_\text{instr}+T_\text{gr}-T_\text{orig})/T_\text{orig}\times 100\%$. \#Event reduction=$(\#ev_\text{total} - \#ev_\text{gr})/\#ev_\text{total}\times 100\%$. Attestation speed E-speed$ = \#ev_\text{total} / (T_\text{instr} + T_\text{gr})$. Data generation speed D-speed$ = Z_s / (T_\text{instr} + T_\text{gr})$.}
\label{tab:event-folding}
\centering
\small


\begin{tabular}{l|r|r|r|r|r|r r|r|r|r|r|r|r r}
\hline
 & \multicolumn{7}{c|}{GCC} & \multicolumn{7}{c}{LLVM} \\
 \cline{2-15}
Program & $T_\text{orig}$ & $T_\text{instr}$ & $T_\text{gr}$ & \#$ev_\text{total}$ & \#$ev_\text{fold}$ & \#$ev_\text{gr}$ & $Zs$ & $T_\text{orig}$ & $T_\text{instr}$ & $T_\text{gr}$ & \#$ev_\text{total}$ & \#$ev_\text{fold}$ & \#$ev_\text{gr}$ & $Zs$ \\
& (s) & (s) & (s) & ($\times 10^3$) & ($\times 10^3$) & ($\times 10^3$) & (KB) & (s) & (s) & (s) & ($\times 10^3$) & ($\times 10^3$) & ($\times 10^3$) & (KB) \\
\hline
400.perlbench & 1.3 & 4.0 & 0.5 & 25,311.0 & 15,471.4 & 15,444.2 & 519.4 & 1.6 & 4.7 & 0.1 & 24,884.0 & 2,855.6 & 2,830.6 & 469.1 \\

401.bzip2 & 10.3 & 12.1 & 0.1 & 205,593.1 & 1,804.5 & 1,742.9 & 566.6 & 11.4 & 13.2 & 0.1 & 205,599.3 & 1,806.7 & 1,745.1 & 566.7 \\

403.gcc & 1.5 & 3.5 & 3.4 & 187,747.3 & 99,408.6 & 97,690.7 & 17,489.3 & 1.5 & 3.3 & 3.5 & 185,831.5 & 100,174.0 & 98,463.0 & 17,579.9 \\

429.mcf & 4.0 & 6.7 & 0.3 & 174,799.9 & 9,767.0 & 7,090.7 & 2,195.7 & 4.4 & 7.0 & 0.3 & 174,799.9 & 9,767.1 & 7,090.7 & 2,241.1 \\

433.milc & 12.0 & 13.7 & 0.0 & 311,950.1 & 15.4 & 15.4 & 3.0 & 16.6 & 18.0 & 0.0 & 313,774.1 & 15.8 & 15.8 & 3.0 \\

445.gobmk & 5.4 & 7.5 & 1.6 & 60,850.8 & 50,976.7 & 50,534.1 & 7,786.2 & 5.2 & 7.4 & 1.6 & 60,859.8 & 50,985.4 & 50,543.0 & 7,781.5 \\

456.hmmer & 7.4 & 8.0 & 0.0 & 79,139.7 & 4.7 & 4.7 & 2.7 & 6.8 & 8.0 & 0.0 & 79,139.7 & 4.7 & 4.7 & 2.7 \\

458.sjeng & 5.6 & N/A & N/A & 383,144.6 & N/A & N/A & N/A & 5.5 & N/A & N/A & 378,466.7 & N/A & N/A & N/A \\

462.libquantum & 0.1 & 0.1 & 0.0 & 1,018.7 & 24.6 & 24.6 & 2.7 & 0.1 & 0.1 & 0.0 & 1,279.3 & 24.7 & 24.7 & 2.6 \\

464.h264ref & 27.9 & 39.6 & 1.3 & 2,059,738.2 & 40,118.8 & 40,032.9 & 2,580.7 & 29.8 & 41.6 & 1.8 & 2,061,382.9 & 52,545.2 & 52,459.3 & 2,976.7 \\

470.lbm$^\text{a}$ & 2.8 & 2.8 & 0.0 & 0.12 & 0.03 & 0.03 & 0.2 & 2.5 & 2.5 & 0.0 & 0.12 & 0.03 & 0.03 & 0.2 \\

482.sphinx3 & 2.1 & 2.3 & 0.0 & 34,596.9 & 842.4 & 728.4 & 166.2 & 2.0 & 2.3 & 0.0 & 34,730.4 & 836.1 & 725.0 & 167.4 \\

\hline
Avg.$^\text{b}$ & \multicolumn{3}{c|}{overhead = 43.7\% } & \multicolumn{4}{c|}{reduction = 93.2\% } & \multicolumn{3}{c|}{overhead = 41.0\%} & \multicolumn{4}{c}{reduction = 93.2\% } \\
\cline{2-15}
& \multicolumn{7}{c|}{E-speed = 29.2M/s\qquad D-speed = 291.3KB/s} & \multicolumn{7}{c}{E-speed = 27.2M/s\qquad D-speed = 275.2KB/s} \\
\hline
\end{tabular}

{\raggedright ~$^\text{a}$ Small numbers of $\#ev$ to two decimal places. \par}
{\raggedright ~$^\text{b}$ \verb|458.sjeng| not taken into account. \par}

\end{table*}

To evaluate the control-flow event folding, we first report the runtime overhead of instrumented program compared with the original program. Then we evaluate the reduction on the control-flow events compared with the case that the control-flow events are unfolded and faithfully recorded. The results are presented in Table~\ref{tab:event-folding}. $T_\text{orig}$ is the execution time of the vanilla program. $T_\text{instr}$ is the execution time of the instrumented program. $T_\text{gr}$ is the time used by the greedy compression algorithm. The overhead of our prover-side approach ranges 0\%$\sim$360\% for GCC binaries and 0\%$\sim$353\% for LLVM binaries. The average time overhead is 42.3\%. For \verb|458.sjeng|, we confront binary instrumentation failure while inserting the code of event folding at specific program-structure-related points.

The key contribution of our approach is to mitigate the explosion of control-flow events. To evaluate this point, we first craft another type of instrumented binaries to only record the number of runtime control-flow events but not fold them. The number of control-flow events recorded by such binaries is represented by $\#ev_\text{total}$ in Table~\ref{tab:event-folding}. Meanwhile, $\#ev_\text{fold}$ is the number of control-flow events derived by our instrumented program. $\#ev_\text{gr}$ is the number of control-flow events after the greedy compression. $Zs$ stands for the amount of data after compressing these events with the Zstandard algorithm, that is the payload of the attestation protocol between the prover and the verifier. The reduction on the control-flow events ranges 17.0\%$\sim$99.9\% for GCC binaries and also 17.0\%$\sim$99.9\% for LLVM binaries. The overall reduction in the control-flow events is 93.2\%.

The time overhead of our approach is considerably more than the traditional local CFI protections (usually $\leq 5\%$). This is because the instrumented operations are more complicated to fold the paths and enable the remote diagnosing on the vulnerable control-flow paths. To address the control-flow events attestation speed, we define \emph{E-speed} as the speed of the prover generating raw runtime control-flow events. As seen in Table~\ref{tab:event-folding}, the peak E-speed is 50.4M/s for GCC binaries and 47.5M/s for LLVM binaries. The average attestation speed is 28.2M/s. Because of the different encodings of the control-flow edges as described in Section~\ref{sec:implementation}, the reduction on control-flow events is different from the reduction on the generated code addresses. The amount of data delivered from the prover to the verifier, as well as the data generation speed, should be of a reasonable magnitude. We define the data generation speed \emph{D-speed} as the speed of the prover generating data that are sent to the verifier. As seen in Table~\ref{tab:event-folding}, the peak D-speed is 2.53MB/s for GCC binaries and 2.59MB/s for LLVM binaries. The average data generation speed is 283.0KB/s. Both the peak and average data generation speeds are adaptive to the wireless network transmission speed for IoT/CPS, which avoid the drastic data accumulation at the prover and justify the usability of our approach. The significant difference between the E-speed and the D-speed also indicates the effectiveness of our optimizations.

\paragraph{RQ.2} How can we decide the proper $BOUND$ for the greedy compression?

In the evaluations in Table~\ref{tab:event-folding}, we set the $BOUND$ value of the sliding window in Algorithm~\ref{algo:greedy} as 4. We get this setting by tuning the value of $BOUND$ on the benchmarks compiled with GCC, as presented in Table~\ref{tab:bound}. The choice of $BOUND$'s value is a spatial-temporal trade-off. A larger $BOUND$ may lead to a higher compression rate with a longer time. Due to the complexity of program control flows, there is no universal optimal value. We use the following formula to help choose the preferable value of $BOUND$ from Table~\ref{tab:bound},
\begin{equation}
max\left(
  Avg_{p\in\textit{benchmarks}}((1-1/\mathcal{R}_p) / T_{\text{gr},p})
  \right)
  \label{eq1}
\end{equation}
For each program, (1-1/$\mathcal{R}$) represents the ratio of compression in another form of $(\#ev_\text{fold}-\#ev_\text{gr})/\#ev_\text{fold}$. Therefore, $(1-1/\mathcal{R})/T_\text{gr}$ measures the gain of compression in each time unit. For each value of $BOUND$, we calculate the average value of this gain for the benchmarks in Table~\ref{tab:bound}. The tuning procedure finally chooses $BOUND=2^2$ used in our evaluations. Intuitively, $T_\text{gr}$ increases exponentially along with the exponential increase of $BOUND$. Meanwhile, the increase in the gain of compression is not exponential thus small $BOUND$ is preferred. However, we may choose other values of $BOUND$ by using optimization functions other than Eq.(\ref{eq1}), considering the tolerance of network capacity and the computing resources of the prover.

\begin{table*}[!ht]
\renewcommand{\arraystretch}{1.1}
\caption{Tuning the Value of $BOUND$. Compression rate $\mathcal{R} = \#ev_\text{fold}/\#ev_\text{gr}$}
\label{tab:bound}
\centering\small

\begin{tabular}{l|c|r|c|r|c|r|c|r}
\hline
 & \multicolumn{8}{c}{$BOUND$} \\
\cline{2-9}
Program & \multicolumn{2}{c|}{$2^2$} & \multicolumn{2}{c|}{$2^3$} & \multicolumn{2}{c|}{$2^4$} & \multicolumn{2}{c}{$2^5$} \\
\cline{2-9}
 & $\mathcal{R}$ & $T_\text{gr}$(s) & $\mathcal{R}$ & $T_\text{gr}$(s) & $\mathcal{R}$ & $T_\text{gr}$(s) & $\mathcal{R}$ & $T_\text{gr}$(s) \\
\hline

400.perlbench & 1.002 & 0.538 & 1.002 & 1.198 & 1.004 & 2.576 & 1.005 & 5.111 \\

401.bzip2 & 1.035 & 0.075 & 1.106 & 0.122 & 1.213 & 0.225 & 1.253 & 0.426 \\

403.gcc & 1.018 & 3.431 & 1.039 & 6.762 & 1.046 & 14.924 & 1.056 & 28.358 \\

429.mcf & 1.377 & 0.309 & 1.470 & 0.517 & 1.488 & 1.112 & 1.492 & 2.197 \\

433.milc & 1.000 & 0.002 & 1.000 & 0.003 & 1.000 & 0.004 & 1.000 & 0.007 \\

445.gobmk & 1.009 & 1.594 & 1.010 & 3.309 & 1.019 & 7.357 & 1.022 & 14.013 \\

456.hmmer & 1.000 & 0.001 & 1.001 & 0.001 & 1.006 & 0.002 & 1.008 & 0.002 \\


462.libquantum & 1.000 & 0.003 & 1.000 & 0.004 & 1.000 & 0.006 & 1.000 & 0.011 \\

464.h264ref & 1.002 & 1.344 & 1.002 & 2.890 & 1.003 & 6.725 & 1.003 & 13.278 \\

470.lbm & 1.000 & 0.001 & 1.000 & 0.001 & 1.000 & 0.001 & 1.000 & 0.001 \\

482.sphinx3 & 1.157 & 0.032 & 1.177 & 0.055 & 1.183 & 0.109 & 1.187 & 0.223 \\
\hline
$Avg$((1-$\frac{1}{\mathcal{R}}$ / $T_\text{gr}$) & \multicolumn{2}{c|}{0.511} & \multicolumn{2}{c|}{0.506} & \multicolumn{2}{c|}{0.496} & \multicolumn{2}{c}{0.492} \\
\hline
\end{tabular}
\end{table*}

\paragraph{RQ.3} What's the effectiveness of the security enforcement of the verifier?

At the verifier, the performance is generally affected by the design of the shadow stack and the scale of the security policy. The key measurements are the size of the two policy mappings $\langle\mathcal{M},\mathcal{F}\rangle$ because the map searches are the critical operations. We present the size of the mappings, i.e., $|\mathcal{F}|$ and $|\mathcal{M}|$, in Table~\ref{tab:map}. Because we use the ordinary workloads on the SPEC2k6 benchmarks, which contain no exploit to hijack the control flows, our verifier can consume all the recovered control-flow events and report the secure execution of the benchmarks.

The total verification time consists of two parts. The first part is the time of the reverse procedure of the greedy compression, i.e. $T_{\text{gr}^{-1}}$ in Table~\ref{tab:map}. This procedure recovers the folded control-flow events collected by the instrumented prover binaries, i.e., $ev_\text{fold}$. The second part is the verification time taking all the control-flow events in $ev_\text{fold}$ as input and enforcing the security policy, i.e., $T_\text{vrf}$ in Table~\ref{tab:map}. The security enforcement is performed on folded control-flow events and requires no recovery of the original runtime control-flow events $ev_\text{total}$. Therefore the control-flow events verification speed is figured out by $\#ev_\text{fold} /(T_{\text{gr}^{-1}} + T_\text{vrf})$, where $\#ev_\text{fold}$ is identical to the statistics in Table~\ref{tab:event-folding}.
The verification speed ranges 13K/s$\sim$1.74M/s for GCC binaries and 26K/s$\sim$1.70M/s for LLVM binaries. The average verification speed is 1.03M/s. In general, our attestation speed (28.2M/s) and verification speed (1.03M/s) is incomparable to the speeds of ScaRR \cite{DBLP:conf/raid/ToffaliniLB0C19} (250$\sim$400K/s for attestation and 1.4$\sim$2.7M/s for verification), because we take different definitions of control-flow event. Our control-flow events include unskippable direct calls, indirect branches, and returns, while ScaRR's control-flow events also take the direct jumps and system calls into account. More potential improvements on the efficiency of our approach are discussed in Section~\ref{sec:discussion}.

\begin{table*}[!t]
\renewcommand{\arraystretch}{1.1}
\caption{Effectiveness of Context-Sensitive Enforcement at Verifier. Verification speed $=\#ev_\text{fold} / (T_{\text{gr}^{-1}} + T_\text{vrf})$}
\label{tab:map}
\centering\small
\begin{tabular}{l|r|r|r|r|r|r|r|r}
\hline
 & \multicolumn{4}{c|}{GCC} & \multicolumn{4}{c}{LLVM} \\
 \cline{2-9}
Program & $|\mathcal{M}|$ & $|\mathcal{F}|$ & $T_{\text{gr}^{-1}}$(s) & $T_\text{vrf}$(s) & $|\mathcal{M}|$ & $|\mathcal{F}|$ & $T_{\text{gr}^{-1}}$(s) & $T_\text{vrf}$(s) \\
\hline
400.perlbench & 4,289 & 15,299 & 0.556 & 18.025 & 4,308 & 15,248 & 0.103 & 6.513 \\

401.bzip2 & 134 & 460 & 0.066 & 0.974 & 129 & 433 & 0.067 & 0.997 \\

403.gcc & 21,879 & 53,159 & 3.455 & 56.417 & 21,740 & 52,417 & 3.527 & 136.505 \\

429.mcf & 5 & 83 & 0.294 & 5.498 & 5 & 84 & 0.292 & 5.658 \\

433.milc & 372 & 1,591 & 0.002 & 0.015 & 372 & 1,618 & 0.001 & 0.016 \\

445.gobmk & 3,191 & 9,969 & 1.646 & 43.629 & 3,184 & 9,986 & 1.644 & 43.828 \\

456.hmmer & 789 & 4,074 & 0.001 & 0.005 & 787 & 4,088 & 0.001 & 0.004 \\

458.sjeng & 273 & 1,247 & N/A & N/A & 273 & 1,367 & N/A & N/A \\

462.libquantum & 234 & 554 & 0.003 & 0.021 & 223 & 560 & 0.003 & 0.021 \\

464.h264ref & 750 & 3,347 & 1.414 & 39.829 & 759 & 3,533 & 1.883 & 50.149 \\

470.lbm & 19 & 74 & 0.002 & 0.000 & 19 & 76 & 0.001 & 0.000 \\

482.sphinx3 & 1,078 & 2,758 & 0.029 & 0.651 & 1,078 & 2,767 & 0.029 & 0.649 \\
\hline
Avg. vrf. speed & \multicolumn{4}{c|}{1.27M/s} & \multicolumn{4}{c}{0.87M/s} \\
\hline
\end{tabular}
\end{table*}

\paragraph{RQ.4} Does our control-flow attestation effectively enforce CFI to resist real-world exploits?

In this section, we evaluate our system to see if our context-sensitive control-flow attestation can detect several real-world code reuse exploits at the prover and if the vulnerable control-flow events can be diagnosed by the remote verifier. We reproduce the exploits presented in Table~\ref{tab:exploits} at the prover. The CVE is also described in \cite{DBLP:conf/ccs/HuQYCHKL18} and the Control Jujutsu proof-of-concept (PoC) exploits \cite{DBLP:conf/ccs/EvansLOSROS15} are reported detectable by the control-flow policy of TypeArmor \cite{DBLP:conf/sp/VeenGCPCRBHAG16}. Because our enforcement of the verifier uses the same security policy as TypeArmor, these PoC attacks are expected to be detected by our approach. After instrumenting the vulnerable binary, we manually examined that the static binary instrumentation does not affect the memory corruption attacks; otherwise, we adjust the overflow data to ensure the attack take effect. The large size of \verb|ffmpeg|'s binary impedes the static instrumentation; thus, we only instrument on a related part of the CFG. Then, we launch the exploit to trigger the vulnerable control-flow edges at runtime and let the binary fold the control-flow events and deliver them to the verifier. Our verifier detects and reports the exact control-flow events that violate the rules.

These exploits corrupt the forward edges therefore the verifier reports violations that some runtime edge target is not in the expected set of targets. For the PoC exploit on Nginx, we detected an invalid control-flow edge from the call site in \verb|ngx_output_chain| to the function \verb|ngx_execute_proc|. For the PoC exploit on Apache httpd, we detected an invalid edge from the indirect call site inside the function \verb|ap_run_dirwalk_stat| to the function \verb|piped_log_spawn|. For the CVE-2016-10190 on ffmpeg, we found an invalid edge issued from the indirect call site inside the function \verb|avio_read|.

\begin{table}[!t]
\renewcommand{\arraystretch}{1.1}
\caption{Real exploits diagnosed by ReCFA}
\label{tab:exploits}
\centering\small
\begin{tabular}{c| c | c| c}
\hline
Program & Source & Type & Detected? \\
\hline

ffmpeg & CVE-2016-10190 & heap corruption & \checkmark \\



Apache httpd & PoC exploit of \cite{DBLP:conf/ccs/EvansLOSROS15} & heap corruption & \checkmark \\

Nginx & PoC exploit of \cite{DBLP:conf/ccs/EvansLOSROS15} & heap corruption & \checkmark \\

\hline
\end{tabular}
\end{table}

\section{Discussion}\label{sec:discussion}

The prototype implementation of ReCFA does not support multi-thread programs. To extend our approach to support multi-thread programs, we may obtain the thread id at each instrumentation point with Dyninst. We label each control-flow event with the thread id. Then the control-flow event sequence for each thread can be compressed and delivered concurrently as separate attestation reports to the verifier. We also have to instrument at the thread start point to tell the verifier when to copy the context of shadow stack for monitoring a new thread. Then at the verifier, we can create monitoring threads corresponding to the prover-side threads to enforce the security policy.

Differently from the local CFI protections that usually merges larger equivalent classes (ECs) of valid targets to improve the efficiency \cite{DBLP:journals/csur/BurowCNLFBP17}, our verifier-side enforcement, especially the mapping $\mathcal{F}$, leverages the original sets of valid targets derived by the state-of-the-art approach, i.e. \cite{DBLP:conf/sp/VeenGCPCRBHAG16}. We believe the throughput of the network communication give us more flexibility to address precision rather than the efficiency at the verifier. However, we argue that larger equivalent classes are still effective in further improving the efficiency of security enforcement. Another choice to improve the efficiency of the map searching is to build a layer of fix-size cache ahead of the map searching for the big mappings, e.g. \verb|403.gcc| and \verb|400.perlbench| of Table~\ref{tab:map}. Due to the locality of control flow, once the set of valid targets of an indirect call site is missed in the cache by a runtime forward edge, we check the hit in the mapping and load the mapping relations of this call site into the cache.

Delivering control-flow events from the prover to the verifier gives us the opportunity to achieve a real-time control-flow attestation compared with the traditional control-flow attestations using the hash digests as path measurements. Though we have not exploited this potential in our implementation, we may further investigate the lagging of the verifier-side anomaly reports concerning the overhead of instrumented binary, the cost of the runtime control-flow event condensing, and the network latency caused by data delivering.

We have not considered the non-control-data attack \cite{DBLP:conf/uss/Chen0S05}, which is claimed to be mitigated by C-FLAT \cite{DBLP:conf/ccs/AberaADENPST16}. In their approach, the valid values for branch-condition data are enumerated to derive the offline measurements, therefore invalid data w.r.t the attack is detected as an abnormal runtime measurement. The prerequisite to enumerate valid values of these data, though feasible for small IoT programs, is difficult to be satisfied by the complicated software.

\section{Related Work}\label{sec:related}

\paragraph{Control-flow attestation}

Compared with the static attestation that attests the integrity of program binaries and configurations, runtime attestation measures the runtime states of the program and the properties of its runtime inputs, outputs, and behaviors \cite{Haldar:2004:SRA:1267242.1267245, DBLP:conf/ccs/DaviSW09, hristozov2018practical, DBLP:conf/ccs/AberaADENPST16, DBLP:conf/iccad/ZeitouniDAS0JS17, DBLP:conf/dac/DessoukyZNPDKAS17, DBLP:conf/raid/ToffaliniLB0C19, DBLP:conf/iccad/DessoukyA0S18, DBLP:journals/corr/abs-2103-12928, DBLP:conf/ndss/AberaBB0SS19, DBLP:conf/trustcom/HuHWWZL19, DBLP:conf/pst/GedenR19, DBLP:conf/css/LiuYLZFL19, DBLP:journals/corr/abs-2011-07400, DBLP:conf/hpcc/HuoWLLWX20, DBLP:conf/sp/SunFLJ20}.

Semantic remote attestation \cite{Haldar:2004:SRA:1267242.1267245} used a trusted language-based virtual machine to attest dynamic properties of platform-independent code running in it to the remote parties. DynIMA \cite{DBLP:conf/ccs/DaviSW09} combined load-time measurements and dynamic taint analysis to enforce the integrity of binaries. The architecture instrumented the program with the tracking code that performs runtime integrity-related checks, e.g. counting the small sequences of instructions between consecutive returns and reporting an ROP attack when finding several small instruction sequences executed consecutively. Several hardware-based runtime attestations resorted to the mechanism on the microcontroller \cite{hristozov2018practical} or specific off-chip hardware security module \cite{DBLP:conf/pst/GedenR19} to generate the runtime attestation evidence.

Control-flow attestation \cite{DBLP:conf/ccs/AberaADENPST16, DBLP:conf/dac/DessoukyZNPDKAS17, DBLP:conf/iccad/ZeitouniDAS0JS17, DBLP:conf/iccad/DessoukyA0S18,DBLP:conf/ndss/AberaBB0SS19, DBLP:conf/css/LiuYLZFL19, DBLP:journals/corr/abs-2011-07400, DBLP:conf/raid/ToffaliniLB0C19, DBLP:conf/trustcom/HuHWWZL19, DBLP:conf/sp/SunFLJ20, DBLP:conf/hpcc/HuoWLLWX20} is a kind of runtime attestation diagnosing the execution path of remote software and ensuring the integrity of the program under control-oriented exploits. C-FLAT\cite{DBLP:conf/ccs/AberaADENPST16} proposed to measure the validity of the execution paths with the CFG of the program being attested. The prover aggregates the feature of executed paths using hash operations to generate cumulative measurements. An unexpected measurement indicates to the verifier that an illegal path has been executed.
Different from the isolated execution environment used by C-FLAT to protect the attestation, LO-FAT\cite{DBLP:conf/dac/DessoukyZNPDKAS17} uses processor features and IP blocks, including branch monitor and hash engine, to implement efficient control-flow attestation for vanilla programs. To address the TOCTOU problem that allows the attack to attest benign code while executing temporary malicious code, ATRIUM \cite{DBLP:conf/iccad/ZeitouniDAS0JS17} attests the executed instructions on independent hardware. DIAT \cite{DBLP:conf/ndss/AberaBB0SS19} adopts control-flow attestation to verify only critical code modules of autonomous systems efficiently.
LiteHAX \cite{DBLP:conf/iccad/DessoukyA0S18} extends the hardware-based attestation scheme to support detecting data-only attacks.
LAPE \cite{DBLP:conf/hpcc/HuoWLLWX20} is a control-flow attestation for bare-metal systems. The instrumented firmware separates the code into attestation compartments. The calls between functions in each compartment are captured as paths at runtime and used to generate the attestation report during compartment switching. MGC-FA \cite{DBLP:conf/trustcom/HuHWWZL19} used a machine-learning model to predict the vulnerable probability and decide the strictness-level of control-flow attestation.
Tiny-CFA \cite{DBLP:journals/corr/abs-2011-07400} provides control-flow attestation for low-end MCUs and only requires the hardware to support PoX architecture \cite{DBLP:conf/uss/NunesERT20}.
OAT \cite{DBLP:conf/sp/SunFLJ20} leverages control-flow attestation to enforce operation-scoped CFI. The measurement was protected by the trusted execution environment and the control flow verification is performed through abstract execution.
Most of these works mainly address the overhead reduction of the control flow verification of small or moderate-size programs on resource-constraint systems.
ScaRR \cite{DBLP:conf/raid/ToffaliniLB0C19} is the first control-flow attestation designed for complex systems. It provides a control-flow model for abstracting execution paths. ScaRR relies on the CFG to generate the measurement database. The path separation between checkpoints can reduce the path explosion on measurements generation but also brings in potential false negatives of the online remote verification caused by the context missing. Besides, ScaRR requires source-code instrumentation to detect and instrument the control-flow events and generate the offline measurements. Contrarily, in this work, we use binary CFG as the security policy and binary-level static analysis for the instrumentation. ReCFA should be used when the completeness of the measurement DB cannot be guaranteed, e.g. for complex programs whose paths and sub-paths are difficult to be completely enumerated in advance. Several control-flow attestation approaches also mitigate non-control-data attacks \cite{DBLP:conf/ccs/AberaADENPST16, DBLP:conf/dac/DessoukyZNPDKAS17, DBLP:conf/iccad/ZeitouniDAS0JS17, DBLP:conf/ndss/AberaBB0SS19, DBLP:conf/trustcom/HuHWWZL19} or data-only attacks \cite{DBLP:conf/iccad/DessoukyA0S18, DBLP:conf/sp/SunFLJ20, DBLP:journals/corr/abs-2103-12928}.

\paragraph{Binary-Level Control-Flow Integrity Protections}

There has been a lot of works on binary-level control-flow integrity (CFI) defenses \cite{DBLP:conf/uss/ZhangS13, DBLP:conf/sp/ZhangWCDSMSZ13, DBLP:conf/ndss/MohanLBHF15, DBLP:conf/dimva/PayerBG15, DBLP:conf/acsac/WangYBSF15, DBLP:conf/sp/VeenGCPCRBHAG16, DBLP:conf/raid/MunteanFTLGE18, DBLP:conf/ndss/KimSZT21}. CCFIR \cite{DBLP:conf/sp/ZhangWCDSMSZ13} enforces coarse-grained CFI policies that allow only a set of white-listed return targets. binCFI \cite{DBLP:conf/uss/ZhangS13} also proposed enforcement of coarse-grained policies on stripped binaries without debugging or relocation information. The constraints on the set of valid targets are relaxed for both backward and forward edges to benefit the performance. O-CFI\cite{DBLP:conf/ndss/MohanLBHF15} uses binary instrumentation to combine coarse-grained CFI with code randomization, which enables the program to resist information disclosure attacks on the targets of control transfers. The integrity check at each indirect branch has an independent valid range of target addresses. \verb|binCC| \cite{DBLP:conf/acsac/WangYBSF15} also restricts the returns using boundary checking. Lockdown \cite{DBLP:conf/dimva/PayerBG15} is a DBI-based CFI protection with a shadow stack. It creates the equivalent class per shared object and profiles the CFG at runtime. The precision of Lockdown relies on the symbol information that is absent in stripped binaries. TypeArmor \cite{DBLP:conf/sp/VeenGCPCRBHAG16} enforces an arity-based forward-edge CFI policy by checking the compatibility of the number of parameters of the callee and the number of arguments supplied at each call-site at the binary level, without considering the type constraints of parameters. $\tau$-CFI \cite{DBLP:conf/raid/MunteanFTLGE18} uses both the types and numbers of function parameters directly extracted from binaries to enforce forward- and backward-edge control flow transfers. BPA \cite{DBLP:conf/ndss/KimSZT21} constructs sound and high-precision binary-level CFG based on the points-to analysis over a new block memory model. BPA provides fine-grained CFI policy efficiently. In general, the state-of-the-art CFI protections focus on local enforcement. None of the above works addresses detecting remote control-flow hijacking or diagnoses the vulnerable trace triggering the exploitation.

\section{Conclusion}\label{sec:conclusion}

Abstracting the runtime control-flow facts and faithfully delivering them to the verifier is the critical issue for the control-flow attestations. To overcome the limitations of the offline path measurement and hash-based attestation on complicated software, we come up with a new control-flow attestation approach to abstract the control-flow facts efficiently and enforce control-flow integrity policy at the binary level. Our prototype implementation, ReCFA, relies on the static binary analysis and binary instrumentation to condense the runtime control-flow facts into deliverable size. ReCFA enforces context-sensitive control-flow integrity with a remote shadow stack and the policy mappings adapting off-the-shelf security policies. Besides, to balance the efficiency and security, ReCFA relies on the kernel-based trust anchor with user-level hardware-assisted memory isolation to protect the prover-side mechanisms. Future work includes supporting multi-thread programs and using stronger hardware features to build a more robust trust anchor. The source code of ReCFA has been made publicly available at \url{https://github.com/suncongxd/ReCFA}.

\begin{acks}

We thank the anonymous reviewers for the constructive comments. We also would like to thank Dr. Xiaozhu Meng for the kind advice in using the instrumentation tool Dyninst. Yumei Zhang, Xinzhi Liu, Cong Sun, and Xiao Kan were supported by the National Natural Science Foundation of China (No. 61872279) and the Key Research and Development Program of Shaanxi (No. 2020GY-004).
\end{acks}

\bibliographystyle{ACM-Reference-Format}
\bibliography{mybibliography}

\end{document}